# Ferroic Domains of Alternating Polar and Nonpolar Orders Regulate Photocurrent in Single Crystalline CH$_3$NH$_3$PbI$_3$ Films Self-grown on FTO/TiO$_2$ Substrate

Boyuan Huang [1,2], Guoli Kong [3], Ehsan Nasr Esfahani [1,2], Shulin Chen [4,8], Qian Li [5], Junxi Yu [1,6], Ningan Xu [7], Ying Zhang [3], Shuhong Xie [6], Haidan Wen [5], Peng Gao [4,9], Jinjin Zhao [3,*], Jiangyu Li [1,2, *]

1. Shenzhen Key Laboratory of Nanobiomechanics, Shenzhen Institutes of Advanced Technology, Chinese Academy of Sciences, Shenzhen 518055, Guangdong, China
2. Department of Mechanical Engineering, University of Washington, Seattle, WA 98195-2600, USA
3. School of Materials Science and Engineering, Shijiazhuang Tiedao University, Shijiazhuang, 050043, China
4. Electron Microscopy Laboratory, and International Center for Quantum Materials, School of Physics, Peking University, Beijing 100871, China
5. X-ray Science Division, Argonne National Laboratory, Lemont, IL 60439, USA
6. Key Laboratory of Low Dimensional Materials and Application Technology of Ministry of Education, and School of Materials Science and Engineering, Xiangtan University, Xiangtan 411105, Hunan, China
7. Oxford Instruments Technology (Shanghai) Co. Ltd., Shanghai 200233, China
8. State Key Laboratory of Advanced Welding and Joining, Harbin Institute of Technology, Harbin 150001, China
9. Center for Nanochemistry, College of Chemistry and Molecular Engineering, Peking University, Beijing 100871, China

The spectacular rise in photovoltaic conversion efficiency (PCE) of halide perovskite solar cells (PSCs) based on CH$_3$NH$_3$PbI$_3$ and related materials has fueled intensive interests in hybrid organic-inorganic perovskites [1–3], yet the very crystalline structure of CH$_3$NH$_3$PbI$_3$ remains ambiguous after extensive researches. While it is generally accepted that its room temperature lattice is tetragonal and thus possesses ferroic ordering, there is heated debate on whether such order is polar or not [4,5], and advocates for either structure can find their supports in X-ray and neutron diffractions [6–9], optic second harmonic generation (SHG) [4,5], macroscopic pyroelectric and ferroelectric measurement [4,10,11], microscopic piezoresponse force microscopy (PFM) [12–15], and density functional theory (DFT) and molecular dynamics (MD) simulations [16–19]. A consequence of a polar order is ferroelectricity and its implications to photovoltaics, yet the ferroelectric nature of CH$_3$NH$_3$PbI$_3$ remains controversial as well [4,5,11], and the correlation between photovoltaic conversion and possible ferroelectricity has not been established except for a number of theoretical studies [16–18]. Here, we present compelling evidences that single crystalline CH$_3$NH$_3$PbI$_3$ films possess ferroic domains with alternating instead of pure polar and nonpolar orders, and that polar domains exhibit reduced photocurrent in contrast to previous

---

theoretical expectations [16–18]. These findings resolve key questions and ambiguity regarding polar nature of CH$_3$NH$_3$PbI$_3$, reconcile the diverse and apparent contradictory data widely reported in the literature, and point a direction toward engineering ferroic domains for enhanced PCE.

CH$_3$NH$_3$PbI$_3$ crystals were self-grown on FTO/TiO$_2$ substrates as reported in our previous study [20] (Figs. S1 and S2 in the Supplementary Information (SI)). Electron backscatter diffraction (EBSD) taken at five different locations of a crystal reveals identical Kikuchi patterns (Fig. S3), indicating its single crystalline nature, which is confirmed by synchrotron X-ray diffraction (XRD) over large surface area on the scale of hundreds of microns, as shown by the profiles of several measured reflections denoted in the pseudo-cubic setting (Fig. S1b). The observed (002)$_c$ reflection in the specular condition suggests that the dominant vertical growth direction of CH$_3$NH$_3$PbI$_3$ crystal is along [001]$_c$, consistent with DFT energetic calculations [21]. Crystallography of CH$_3$NH$_3$PbI$_3$ was further examined by high-resolution transmission electron microscopy (HRTEM), revealing well-ordered crystalline lattice (Fig. S1c), and the corresponding electron diffraction pattern in the inset is identified to be tetragonal with viewing direction along the [110]$_t$ or [001]$_t$ zone axis, which are equivalent in pseudo-cubic setting [22] and consistent with XRD observation.

We first present unambiguous ferroic domain structures widely observed in our crystals (Figs. S4, S5), which is broadly aligned with recent literatures [13–15,23], yet on larger scales with richer varieties. This is evident from domains revealed by SEM, AFM topography, polarized optic microscopy and PFM (Fig. 1), all exhibiting characteristic lamellar patterns often seen in ferroics [24–27]. Striking PFM amplitude mapping acquired through single-frequency scanning is shown in Fig. 1d, and the alternating domains exhibit large contrast in piezoresponses up to one order of magnitude difference, hinting their different origins. Tetragonal CH$_3$NH$_3$PbI$_3$ has three ferroic variants with equivalent transformation strains [27], regardless of it being polar I4cm or non-polar I4/mcm. Without loss of generality, we consider four possible domain walls between variant 3 and variants (1, 2), which form different angles when these walls intersect different crystalline facets [27]. Some of the expected angles on selected crystalline planes are summarized in Table S1, which are indeed observed in our experimental domains patterns (Fig. S6). The most common angle between domain walls is 70.5°, suggesting that the corresponding crystalline facet is (101)$_t$ in a tetragonal lattice, consistent with [001]$_c$ growth direction observed in XRD. Furthermore, we also observed 60º expected on (111), 58.5º on (112), and 83.6º expected on (201) planes, and Euler angles in the corresponding ranges are also found in EBSD (Fig. S7). Some of the earlier studies attributed the PFM domains observed in CH$_3$NH$_3$PbI$_3$ to be ferroelectric [15] while others attributed them to be ferroelastic instead [13,14], and such contradictory data exist widely in the literature (Table S2), compelling us to examine in details the nature of ferroic domains of CH$_3$NH$_3$PbI$_3$.

The very presence of a sharp piezoresponse contrast in unambiguous ferroic domains suggests two distinct mechanisms in high- and low-response domains, and we argue that one is polar with true piezoelectricity, and the other is nonpolar with apparent yet weaker piezoresponse arising from electrochemical ionic activities [28–30]. The first evidence lies in lateral piezoresponse signal, one of the key differentiators of true piezoelectricity [31]. Characteristic lateral PFM mapping is indeed observed in our crystals (Fig. S8a) obtained via single-frequency scan, and almost identical domain pattern also appears in vertical PFM (Fig. S8b), exhibiting simultaneously high (or low) lateral and vertical piezoresponses. For a crystal in 4mm point group that a polar tetragonal CH$_3$NH$_3$PbI$_3$ belongs to, vertical piezoresponse reflects contribution from piezoelectric coefficient $d_{33}$ with out-of-plane polar axis, while lateral piezoresponse from piezoelectric coefficient $d_{15}$ with polar axis in-plane [32]. Thus domains with simultaneously high vertical and lateral piezoresponses can arise from [001]$_t$ polar axis inclined from the crystal surface, and the measured domain wall angle of 83.6º (Fig. S6) suggests that the inclination angle is 63.4º (Table S1). However, domains with simultaneous low vertical and lateral piezoresponses in general is not compatible with 4mm point group symmetry, suggesting that they are actually nonpolar. To confirm this interpretation, we carry out lateral PFM scan using dual amplitude resonance tracking (DART) corrected via simple harmonic oscillator (SHO) model [33]. Remarkably, SHO works well for lateral PFM in high-response polar domains but completely fails in low-response nonpolar domains, as demonstrated by black dots for all the failing points in the corrected amplitude mapping (Fig. 2a) that completely cover the low-response domains (Fig. S9a). Mechanisms other than piezoelectricity such as electrostatic interactions and ionic activities are highly unlikely to induce lateral piezoresponse because of their high symmetry [31].

A separate vertical DART PFM scan (Fig. 2b) provides strong support for alternating polar and nonpolar domains in their drastically different energy dissipation, wherein resonance tracking works well according to frequency mapping (Fig. 2c). Additional DART PFM mappings are shown in Fig. S10. It is observed that low-response nonpolar domains exhibit much lower quality factor than high-response polar ones (Fig. 2d), and thus much higher energy dissipation [33]. This is a solid proof that low- and high-response domains do have distinct microscopic mechanisms, and the dissipative ionic processes often observed in $CH_3NH_3PbI_3$ [28–30] are known to result in apparent (though usually much weaker) vertical piezoresponse. Further proof of the piezoelectricity in polar domains and nonpiezoelectric mechanisms in nonpolar ones are revealed by their distinct first and second harmonic electromechanical responses, probed as schematically shown in Fig. S11. It is observed that first harmonic linear response dominates second harmonic quadratic ones in high-response polar domains (Fig. 2e), while the trend is reversed for low-response nonpolar domains, suggesting that the former is piezoelectric and the latter is not [31]. Similar observations have been made throughout these two types of domains and at different AC excitations (Fig. 2f). In addition, the corresponding variations of quality factor in Fig. S12 reveal lower quality factor and thus higher dissipation in low-response nonpolar domains, as expected. Importantly, for high-response polar domains, quality factors in first harmonic response (originating from true piezoelectricity) are higher than those of second harmonic ones (originating from nonpiezoelectric dissipative processes), indicating two distinct microscopic mechanisms (Fig. 2e). Low-response nonpolar domains, on the other hand, have similar quality factors for both first and second harmonic responses at all voltages, indicating that they both arise from nonpiezoelectric dissipative processes. We have also acquired mappings of first and second harmonic piezoresponses in a different domain pattern (Fig. 2g), revealing that domains with high (low) first harmonic response (Fig. 2h) exhibit low (high) second harmonic response (Fig. 2i), confirming without ambiguity the opposite trends and distinct microscopic mechanisms in these polar and nonpolar domains.

The alternating polar and nonpolar ferroic domains in $CH_3NH_3PbI_3$ also exhibit opposite temperature evolutions across phase transition. It is found that a thermal probe heated by 3.5V can largely erase the AFM topography domains (Fig. 3a), which reappears after the heating voltage is reduced to 1V to decrease the temperature. This suggests reversible cubic-to-tetragonal phase transition induced by heating and cooling, as recently reported in twins from TEM observation [23]. Furthermore, mappings of the topography (Fig. 3b) and PFM (Fig. 3c) under global heating reveal that contrast in lamellar ferroic domains seen at 30°C and 35°C starts to decrease with increased temperature, and they disappear altogether at nominal temperature of 67°C and 70°C. Remarkably, domain structures reemerge in both topography and PFM upon cooling, fully recovering at 35°C, and similar phenomena are observed in two cycles of heating and cooling (Fig. S13), indicating a two-way memory effect in $CH_3NH_3PbI_3$ crystal. Such memory effect is better visualized by the line scan of topography before and after heating (Fig. 3d), wherein clear roof-like topography feature seen at 30°C largely disappears at 67°C, and then remerges at 35°C with identical inclination. Much insight is gained from the evolution of piezoresponse with respect to temperature across phase transition (Fig. 3e), where it is observed that piezoresponse increases upon heating in low-response nonpolar domains yet decreases in high-response polar domains, consistent with expected piezoelectric and ionic mechanisms [34], and they converge above phase transition, since nonpiezoelectric contributions remain in high-response domains beyond phase transition when the crystal becomes completely nonpolar.

Finally, we show that ferroic domains correlate with photocurrent as revealed by simultaneous mappings of PFM, photoconductive AFM (pcAFM) and Kelvin probe force microscopy (KPFM). The PFM amplitude in Fig. 4a shows characteristic lamellar domain pattern, which is followed closely by photocurrent distributions [15] mapped in the same area without a DC bias (Fig. 4b). Furthermore, it is observed that photocurrent in high response polar domains is smaller, suggesting that polar order reduces photocurrent in $CH_3NH_3PbI_3$ on $FTO/TiO_2$. Such correlation between ferroic domains and electric characteristics is further supported by KPFM acquired in the same area as Fig. 3(a-d), revealing a negative shift of surface potential in polar domains (Fig. 4c), consistent with reduced photocurrent observed due to electron-collecting nature of $FTO/TiO_2$ substrate. Similar to AFM topography and PFM amplitude upon heating and cooling across phase transition, memory effect is also observed in photocurrent distribution (Fig. 4d-f). The domain contrast in photocurrent seen at 35°C largely disappears after heating to 70°C, yet reemerges after cooling to 35°C. This further demonstrates that photocurrent of $CH_3NH_3PbI_3$ crystal on $FTO/TiO_2$ substrates is regulated by ferroic orders.

Alternating polar and nonpolar structures in ferroic domains resolves a key puzzle of $CH_3NH_3PbI_3$, that they exhibit strong piezoresponse in polar domains, yet cannot be switched by an electric field applied either locally through a SPM tip or globally through external electrodes [13,14]. We believe this is precisely the consequence of alternating polar and nonpolar domains, since polar domains that are normally switchable by an electric field are now locked by nonpolar ones that sandwich them. DFT calculations consistently predict that polar structure of tetragonal $CH_3NH_3PbI_3$ is energetically favored at room temperature, though the difference is rather subtle, only in the order of 10s meV [16–18]. Thus such alternating polar and non-polar domains can be understood from energy landscape predicted from DFT, and such subtle differences on a fine spatial scale are extremely difficult to distinguish. Scanning probes, on the other hand, are powerful in resolving their distinct functional responses with high spatial resolution. Importantly, $FTO/TiO_2$ substrates are widely used in PSCs, and thus reduced photocurrent observed in polar $CH_3NH_3PbI_3$ on $FTO/TiO_2$ is significant for devices. Whether such reduction is caused by the intrinsic photovoltaic properties of polar domains or arises from structural effects at interface between $CH_3NH_3PbI_3$ and $FTO/TiO_2$, for example by band bending due to polarization, requires further investigations. Nevertheless, our studies resolve otherwise indistinguishable polar and nonpolar domains in $CH_3NH_3PbI_3$, and point a direction toward engineering ferroic domains for enhanced PCE.

**Acknowledgement**

We acknowledge National Key Research and Development Program of China (2016YFA0201001, 2016YFA0300804), US National Science Foundation (CBET-1435968), National Natural Science Foundation of China (11627801, 11472236, 11772207, 51502007, and 51672007), the Leading Talents Program of Guangdong Province (2016LJ06C372), and Shenzhen Knowledge Innovation Program (JCYJ20170307165905513). This material is based in part upon work supported by the State of Washington through the University of Washington Clean Energy Institute. This research used resources of the Advanced Photon Source, a U.S. Department of Energy (DOE) Office of Science User Facility operated for the DOE Office of Science by Argonne National Laboratory under Contract No. DE-AC02-06CH11357. We also acknowledge Electron Microscopy Laboratory in Peking University for the use of aberration corrected transmission electron microscope. J.L. and J.Z. conceived and supervised the project; J.Z., G.K. and Y.Z. grew the crystals and did SEM; B.H. carried out AFM and optic microscopy studies with assistances of E.E., J.Y., and S.X.; J.L. carried out crystallographic analysis of ferroic domains; S.C. and P.G. carried out and analyzed HRTEM; Q.L. and H.W. carried out and analyzed synchrotron XRD; A.X., J.Z. and G.K. carried out and analyzed EBSD. J.L. wrote the manuscript, and all authors participated in the revision.

**Figure Captions**

**Fig. 1** Ferroic domain patterns of $CH_3NH_3PbI_3$ crystals revealed by (a) SEM; (b) AFM topography; (c) polarized optic microscopy; and (d) vertical PFM amplitude.

**Fig. 2** Alternating polar and nonpolar domains in $CH_3NH_3PbI_3$ crystal; (a) lateral PFM mapping showing complete failure of SHO (marked by black dots) in nonpolar domains due to their lack of true piezoelectricity; (b) vertical PFM mappings showing identical domain pattern consisting of high-response polar domains and low-response nonpolar domains; (c) resonant frequency mapping of vertical PFM showing elastic contrast between polar and nonpolar domains and good resonance tracking; (d) quality factor mapping showing substantially lower quality factor and thus higher energy dissipation in nonpolar domains; (e) point-wise tuning of piezoresponse versus frequency showing a point in high-response polar domain has dominant first harmonic response and negligible second harmonic one, while a point in low-response nonpolar domain has higher second harmonic response; (f) comparison of first and second harmonic responses versus AC voltages averaged over a number of points in high- and low-response domains confirming the trend in (e); and (g) AFM topography domain pattern with (h) first and (i) second harmonic mappings confirming opposite contrast of first and second harmonic response.

**Fig. 3** Opposite temperature variations in polar and nonpolar domains of $CH_3NH_3PbI_3$ crystal across phase transition; (a) erasing and reappearing of AFM topography domains by heating and cooling using a heated thermal probe; (b) AFM topography and (c) PFM mappings under a sequence of temperature across phase transition showing appearance and reemergence of ferroic domains; (d) topography line scan indicated in (b) before and after heating showing the disappeared topography feature at high temperature is fully recovered upon cooling; (e) piezoresponses averaged in high-response polar and low-response nonpolar domains showing opposite trend with respect to temperature, yet convergence beyond phase transition.

**Fig. 4** Correlation between photocurrent and ferroic domains of $CH_3NH_3PbI_3$ crystal across phase transition; (a) PFM mapping; (b) photocurrent distribution under no DC bias following ferroic domain pattern in (a) with reduced photocurrent in polar domains; (c) surface potential distribution under light follows ferroic domain pattern in Fig. 3(a-d) with negatively shifted potential in polar domains; and (d) photocurrent distribution in a separate domain pattern at different temperatures across phase transition, showing the disappearing domain pattern at 70°C upon heating and its reemergence at 35°C after cooling.

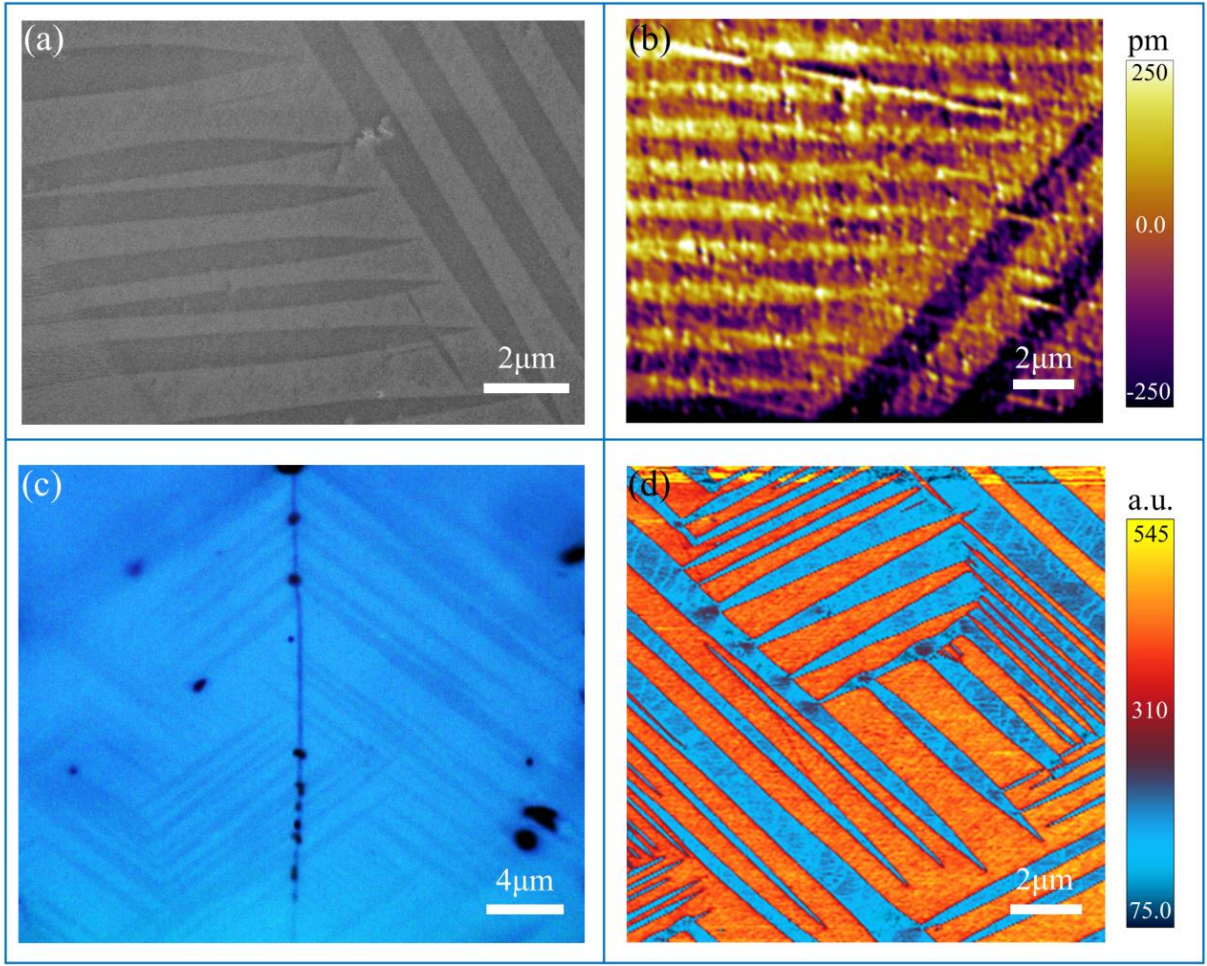

**Fig. 1** Ferroic domain patterns of CH$_3$NH$_3$PbI$_3$ crystals revealed by (a) SEM; (b) AFM topography; (c) polarized optic microscopy; and (d) vertical PFM amplitude.

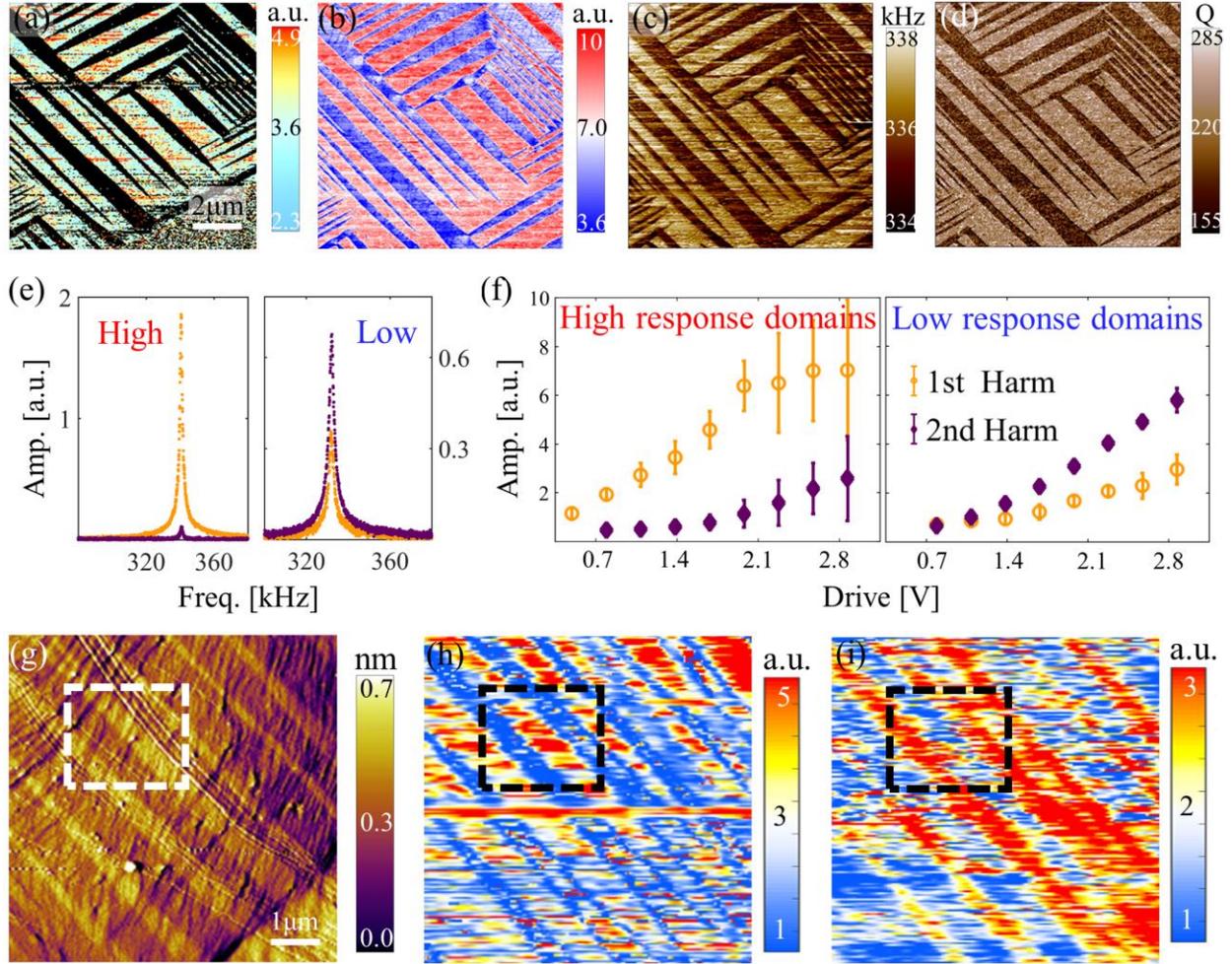

**Fig. 2** Alternating polar and nonpolar domains in $CH_3NH_3PbI_3$ crystal; (a) lateral PFM mapping showing complete failure of SHO (marked by black dots) in nonpolar domains due to their lack of true piezoelectricity; (b) vertical PFM mappings showing identical domain pattern consisting of high-response polar domains and low-response nonpolar domains; (c) resonant frequency mapping of vertical PFM showing elastic contrast between polar and nonpolar domains and good resonance tracking; (d) quality factor mapping showing substantially lower quality factor and thus higher energy dissipation in nonpolar domains; (e) point-wise tuning of piezoresponse versus frequency showing a point in high-response polar domain has dominant first harmonic response and negligible second harmonic one, while a point in low-response nonpolar domain has higher second harmonic response; (f) comparison of first and second harmonic responses versus AC voltages averaged over a number of points in high- and low-response domains confirming the trend in (e); and (g) AFM topography domain pattern with (h) first and (i) second harmonic mappings confirming opposite contrast of first and second harmonic response.

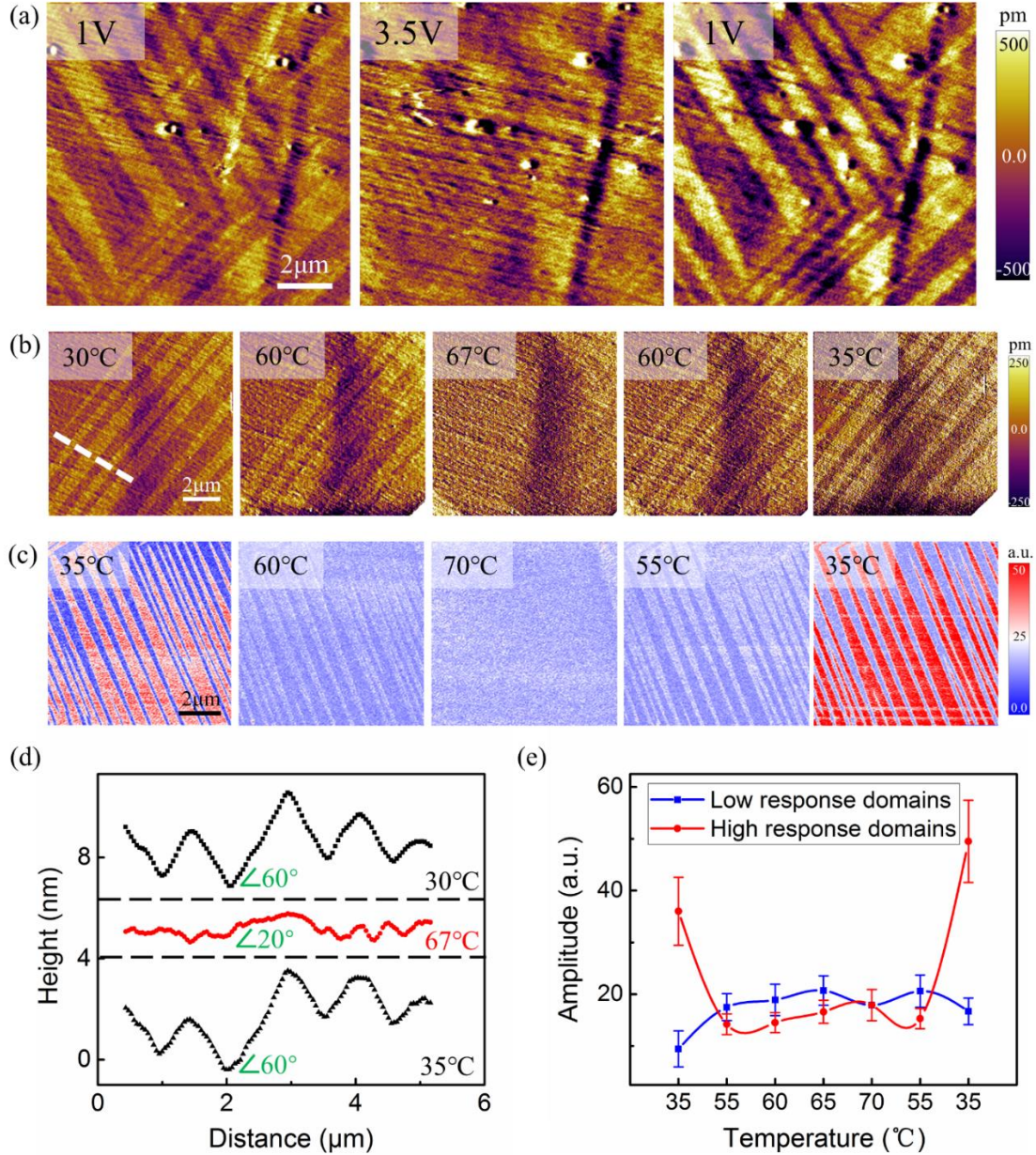

**Fig. 3** Opposite temperature variations in polar and nonpolar domains of $CH_3NH_3PbI_3$ crystal across phase transition; (a) erasing and reappearing of AFM topography domains by heating and cooling using a heated thermal probe; (b) AFM topography and (c) PFM mappings under a sequence of temperature across phase transition showing appearance and reemergence of ferroic domains; (d) topography line scan indicated in (b) before and after heating showing the disappeared topography feature at high temperature is fully recovered upon cooling; (e) piezoresponses averaged in high-response polar and low-response nonpolar domains showing opposite trend with respect to temperature, yet convergence beyond phase transition.

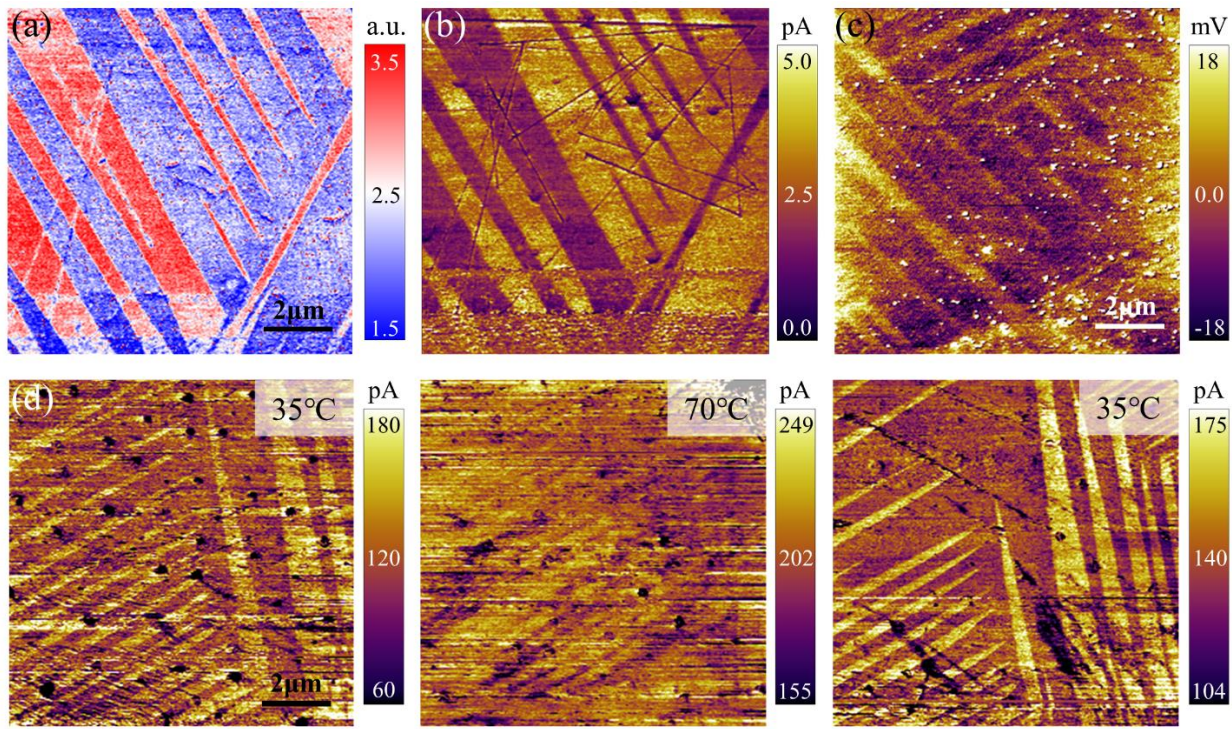

**Fig. 4** Correlation between photocurrent and ferroic domains of $CH_3NH_3PbI_3$ crystal across phase transition; (a) PFM mapping; (b) photocurrent distribution under no DC bias following ferroic domain pattern in (a) with reduced photocurrent in polar domains; (c) surface potential distribution under light follows ferroic domain pattern in Fig. 2(a-d) with negatively shifted potential in polar domains; and (d) photocurrent distribution in a separate domain pattern at different temperatures across phase transition, showing the disappearing domain pattern at 70°C upon heating and its reemergence at 35°C after cooling.

# Ferroic Domains of Alternating Polar and Nonpolar Orders Regulate Photocurrent in Single Crystalline $CH_3NH_3PbI_3$ Films Self-grown on $FTO/TiO_2$ Substrate

Boyuan Huang, Guoli Kong, Ehsan Nasr Esfahani, Shulin Chen, Qian Li, Junxi Yu, Ningan Xu, Ying Zhang, Shuhong Xie, Haidan Wen, Peng Gao, Jinjin Zhao, Jiangyu Li

## Supplementary Information

**Materials and Methods**

**Crystal growth**: Single crystalline $CH_3NH_3PbI_3$ was synthesized following a process reported earlier (*20*). The morphologies of $CH_3NH_3PbI_3$ crystals were examined by scanning electron microscopy (SEM, ZEISS GeminiSEM 300) and atomic force microscopy (AFM, Asylum Research Cypher ES). Electron backscatter diffraction (EBSD) Kikuchi patterns were obtained on Oxford NordlysMax3 detector and analyzed using AZtecHKL software.

**XRD**: Single crystal X-ray diffraction was performed at the Sector 7-ID-C beamline of the Advanced Photon Source, Argonne National Laboratory. The primary X-ray beam with an energy of 10 keV was selected from a diamond-(111) double crystal monochromator with its size defined by a 300×300 $\mu m^2$ slit. A Huber six-circle diffractometer coupled with a PILATUS 100K area detector was employed for alignment of the single-crystalline samples and measurement of both specular and off-specular reflections. The detector was placed downstream from the samples such that ~8° coverage in the 2θ-angle and ~3° in the χ-angle was obtained. Rocking (ω-angle) scans around each reflection were recorded and typically, resultant 3D data volume was reduced by the χ-projection for data analysis.

**HRTEM**: HRTEM and selected area electron diffraction (SAED) were acquired at an aberration-corrected TEM (Titan Cubed Themis G2, FEI) operated at 80 kV. The TEM samples were prepared in an argon-filled glovebox to avoid side reactions. The samples were firstly scratched from substrate and dispersed into anhydrous ether with manual shake for about 2 minutes. Then, the clear suspensions were deposited on holey carbon copper girds for TEM characterizations. The holey carbon copper girds was sealed with a plastic bag full of argon and then transformed into the TEM column.

**Optical microscopy**: Polarized optical microscopy were carried out in reflection-mode using a commercial Nikon Eclipse 80i microscope with a Lumenera Infinity 1-3C camera. The sample was fixed and illuminated by Nikon LV-UEPI2 universal EPI illuminator (a LV-HI50W 12V-LI Halogen Lamp) through top objective lens. The domain pattern was observed by rotating the inside polarizer at an angle $\theta$ with respect to the x-axis of images.

**Atomic force microscopies**:

AFM-based measurements, including topography, PFM, KPFM, and pcAFM, were performed on Asylum Research Cypher AFM and MFP-3D Bio AFM. Nanosensors PPP-EFM conductive probes with a spring constant around 2.8 $Nm^{-1}$ were used for most measurements. Olympus AC160TS-R3 probes with resonance frequency of 288 kHz were also used.

Topography: For topography mapping, the less invasive tapping mode was adopted to avoid possible damage to samples. The probe was excited at its resonance frequency and the scan rate is set to be about 1.0 Hz for visualizing the domain structure. The amplitude channel was used to present topography due to its higher sensitivity to the topography variance. Topography mappings were also acquired from the simultaneous deflection channel of PFM and pcAFM.

PFM: For PFM measurements, we employed four different techniques. The first is single frequency PFM scanning, which provided qualitative contrast between domains with a fast scan rate between 1 and 2 Hz. The vertical and lateral PFM responses were detected by using vertical and lateral deflection signals of cantilever. An AC voltage of 2 V was applied to the probe near the sample–probe resonance frequency $f_0$ to enhance the sensitivity. The $f_0$ of the first and second mode of vertical PFM are 357 kHz and 1079 kHz, separately. The $f_0$ of lateral PFM is 658 kHz.

The second is the dual-amplitude resonance tracking (DART) PFM scanning, which minimizes crosstalk from topography. We used an AC amplitude of 1 V for both vertical and lateral PFM and a slow scan rate below 0.6 Hz to ensure resonant frequency tracking. The mappings of corrected amplitude, phase, resonant frequency, and quality factor were then calculated by using SHO model.

The third is a home-developed first and second harmonic resonance tracking PFM scanning, which was implemented using a Zurich Instrument HF2LI lock-in amplifier in combination with Cypher AFM. In order to obtain the second harmonic signals, the sample response was measured at $f_0$ while the excitation voltage was applied at $f_0/2$. The PID controller inside the amplifier was used to tracking resonance frequency. All data processing was based on SHO model implemented in MATLAB.

The fourth is the point-wise first and second harmonic PFM tunings, which eliminate crosstalk with topography. For each single point, a series of AC voltages from 0.5 to 2.9 V were applied to the probe with an increment of 0.3 V. At each voltage step, the probe-sample system is excited around $f_0$ first and then around $f_0/2$, thus generating two set of tuning data around $f_0$. The corresponding first and second harmonic responses, including amplitude, quality factor, and resonance frequency, were extracted by fitting the raw data with the SHO model.

pcAFM: pcAFM mapping is based on the cAFM mode of AFM using Asylum Research ORCA module, which includes a transimpedance of an amplifier with a gain value of $5\times10^8$ volts per amp. The samples were illuminated by two different sources depending on AFM. For Cypher, the built-in LED sheds light on the sample from the top. For MFP-3D Bio, the illumination was from below through a glass fiber connected to a Nikon C-HGFIE illuminator with an ultrahigh pressure 130W mercury lamp inside. A small DC voltage were applied to the FTO substrate when necessary.

KPFM: For KPFM measurements, the probe scanned the surface topography using tapping mode first and then a 1 V AC voltage was applied on the probe near its resonance frequency to measure the sample surface potential distribution through a DC voltage feedback loop. All KPFM mappings were performed at room temperature with a lift height of -30 nm. The samples were exposed to light illumination if necessary.

Local heating: For local phase transition measurement, the nanoscale area was heated via an Anasys Instruments AN2-300 thermal probe with a resistance about 1 kΩ on the MFP-3D Bio AFM. Temperature can be roughly regulated by changing the amplitude of AC voltages applied to the probe. The method was based on contact mode so that clear topography mapping can be collected from deflection signal.

Global heating: All larger scale phase transition measurements were conducted by heating up the whole airtight sample chamber controlled by an environment controller inside Cypher AFM. A feedback loop can maintain the set temperature with a precision of 0.01°C, which may lead to some noises for PFM mapping due to thermal expansion. The sample temperature was not calibrated due to the tight constraint of the chamber. After setting the target temperature each time, we waited for at least 1 min before any measurement so that there is enough time for thermal equilibrium.

**Additional Data**

**<u>SEM, TEM, XRD, and EBSD</u>:**

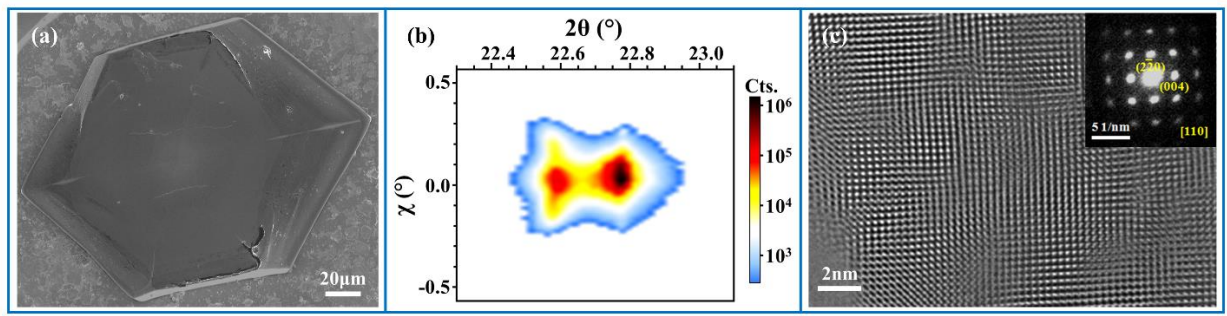

**Fig. 1** Single crystalline $CH_3NH_3PbI_3$ directly grown on $FTO/TiO_2$ substrates; (a) top view SEM image; (b) selected reflections from synchrotron XRD integrated over a rocking angle range of 1 degree; and (c) HRTEM image with the corresponding selected area electron diffraction pattern.

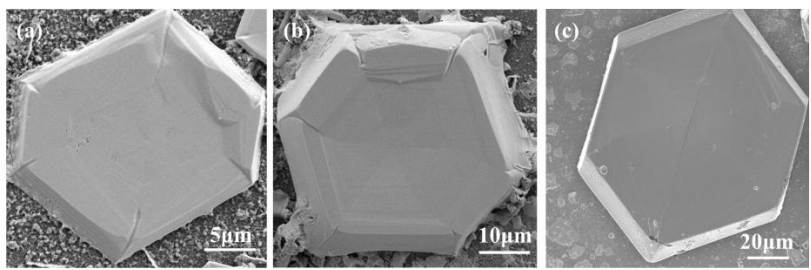

**Fig. S2** SEM images of additional $CH_3NH_3PbI_3$ hexagons.

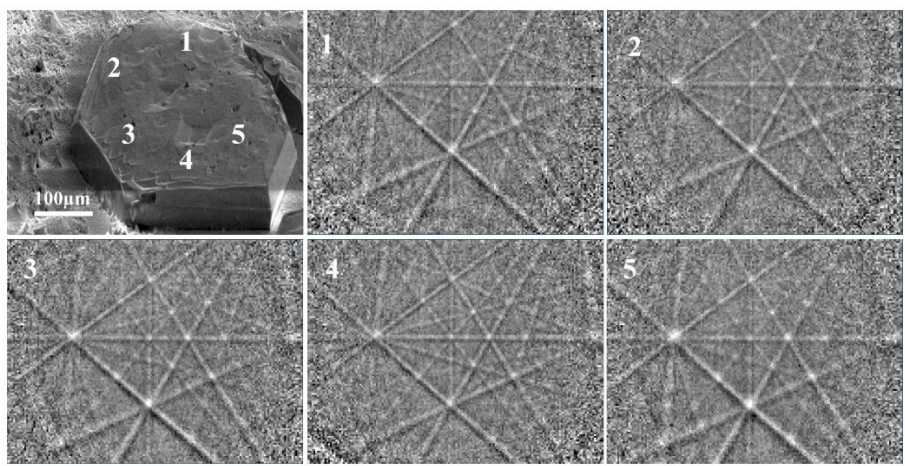

**Fig. S3** EBSD data of a $CH_3NH_3PbI_3$ hexagon reveal identical Kikuchi patterns on five different locations, suggesting that it is a single crystal.

**Ferroic domain patterns:**

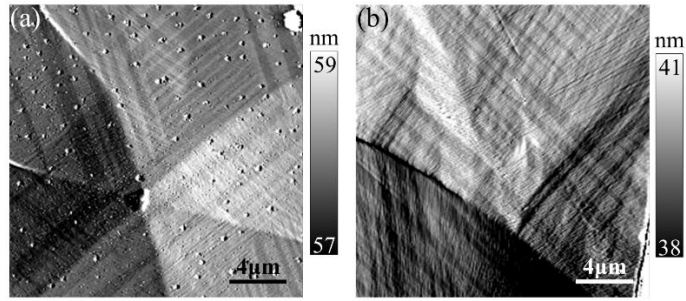

**Fig. S4** Large scale domain patterns in CH$_3$NH$_3$PbI$_3$ crystals revealed by AFM topography.

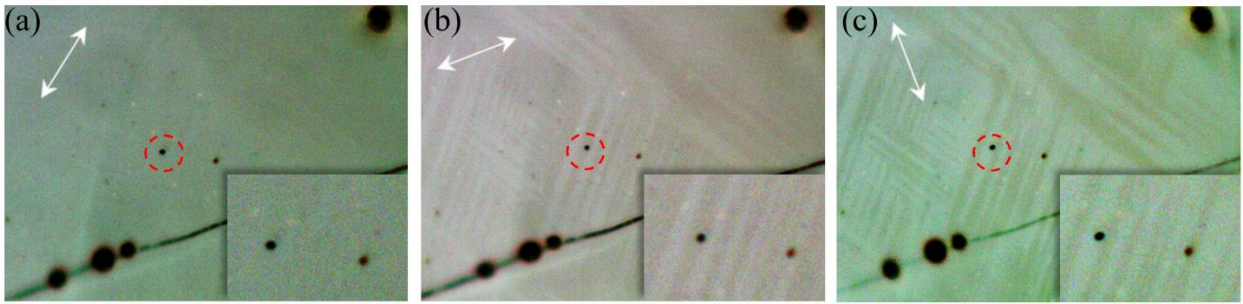

**Fig. S5** Polarized optic microscopy images showing alternating contrast of CH$_3$NH$_3$PbI$_3$ crystal when the polarizer is rotated.

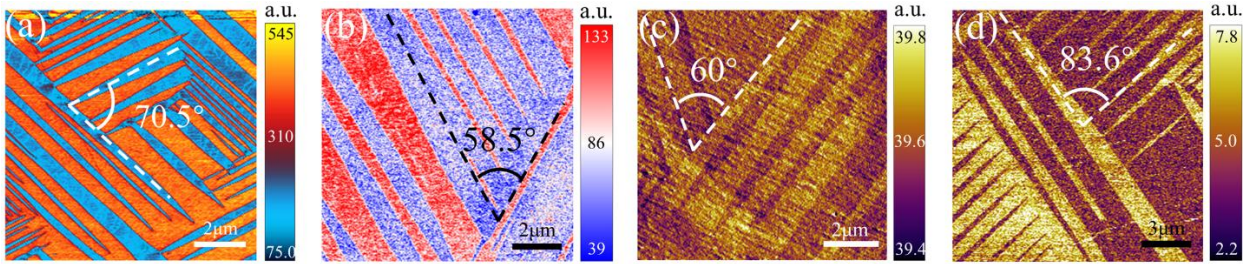

**Fig. S6** Domain patterns of CH$_3$NH$_3$PbI$_3$ crystal with angles between domain walls marked.

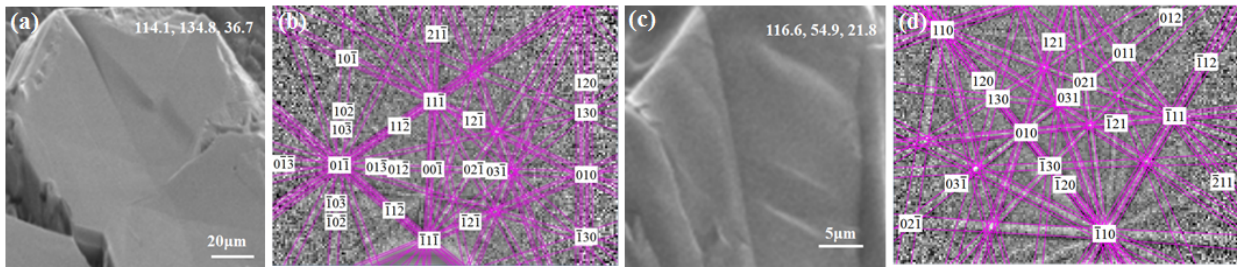

**Fig. S7** EBSD data with Euler angles corresponding to the identified crystalline surfaces

**Polar and nonpolar domain patterns with distinct characteristics:**

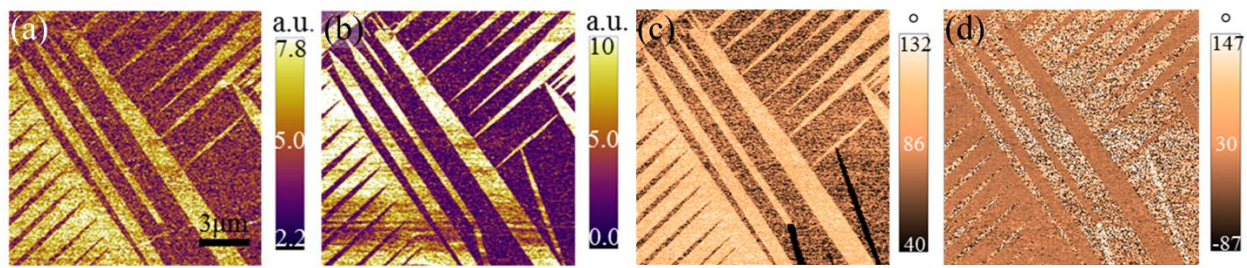

**Fig. S8** Mappings of (ac) lateral and (bd) vertical PFM amplitude (ab) and phase (cd).

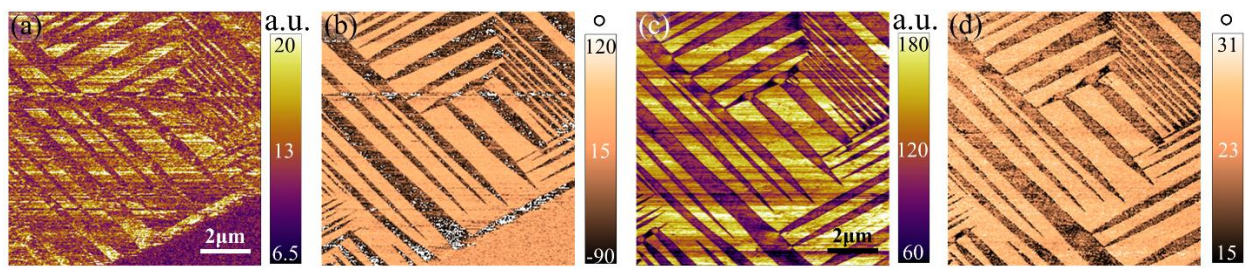

**Fig. S9** Uncorrected (ab) lateral and (cd) vertical PFM amplitude (ac) and phase mapping (bd).

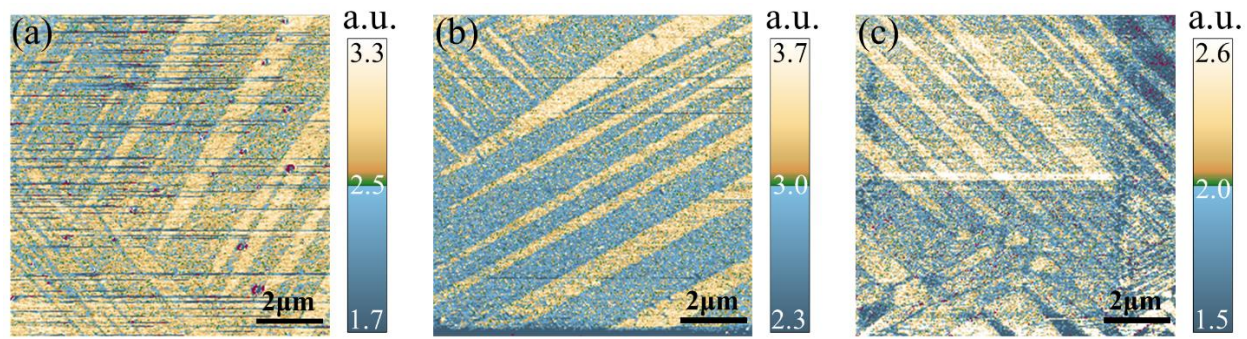

**Fig. S10** Additional vertical PFM amplitude mappings for different domains from DART.

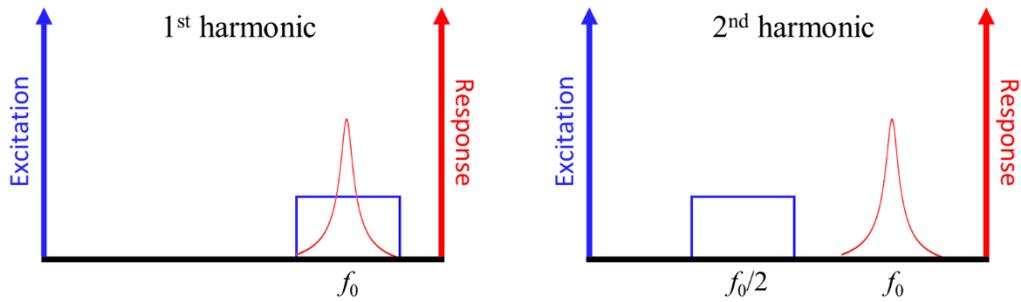

**Fig. S11** Schematics of first and second harmonic piezoresponse measurements.

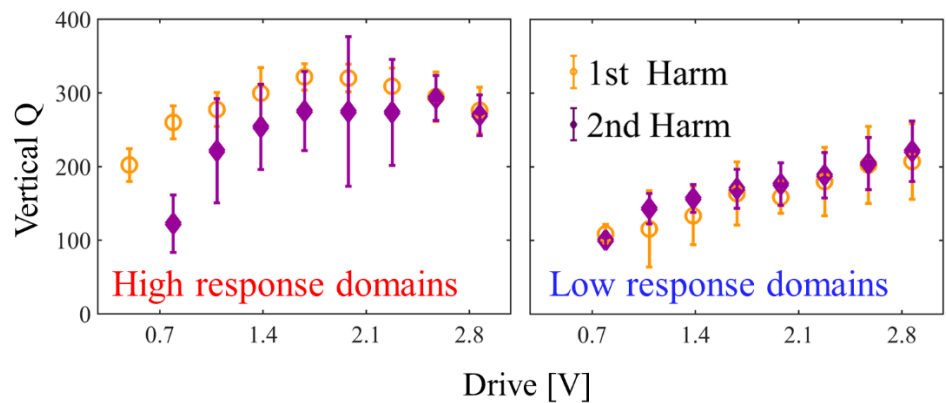

**Fig. S12** Variation of first and second harmonic quality factors versus AC voltage in polar and nonpolar domains.

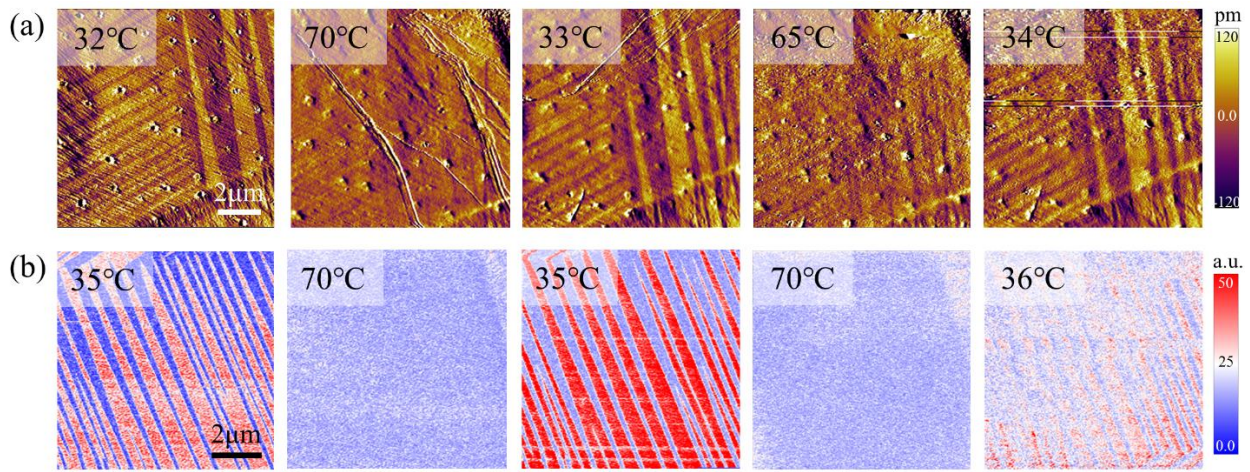

**Fig. S13** Mappings of (a) AFM topography and (b) PFM amplitude under two thermal cycles.

**Table S1** Transformation strains, domain walls, crystal surfaces, and angles between walls on the crystal surfaces, all expressed in a cubic coordinate system that is parallel to the tetragonal crystalline lattice. Values highlighted in bold are found in our experiments as shown in Fig. S4.

| | Variant 3 | Variant 1 | | Variant 2 | |
|---|---|---|---|---|---|
| | $\begin{bmatrix} \alpha & & \\ & \alpha & \\ & & \beta \end{bmatrix}$ | $\begin{bmatrix} \beta & & \\ & \alpha & \\ & & \alpha \end{bmatrix}$ | | $\begin{bmatrix} \alpha & & \\ & \beta & \\ & & \alpha \end{bmatrix}$ | |
| Domain walls between variant 3 and variant 1 or 2 | | (101) | ($10\bar{1}$) | (011) | ($01\bar{1}$) |
| Different crystalline surface | Intersection of domain walls with crystalline surface and the angles between intersecting lines. | | | | |
| | **(101)** surface, Euler angle θ=**45º** | - | [010] | [$11\bar{1}$] | [$\bar{1}11$] |
| | | 54.7º between domain walls (010) & ($11\bar{1}$) or ($\bar{1}11$), **70.5º** between domain walls ($11\bar{1}$) & ($\bar{1}11$) | | | |
| | **(111)** surface, Euler angle θ=**54.7º** | [$10\bar{1}$] | [$1\bar{2}1$] | [$01\bar{1}$] | [$\bar{2}11$] |
| | | 90º between domain walls [$10\bar{1}$] & [$1\bar{2}1$], or [$01\bar{1}$] & [$\bar{2}11$], **60º** between domain walls [$10\bar{1}$] & [$01\bar{1}$], or [$1\bar{2}1$] & [$\bar{2}11$] 30º between domain walls [$10\bar{1}$] & [$\bar{2}11$], or [$1\bar{2}1$] & [$01\bar{1}$] | | | |
| | **(112)** surface, Euler angle θ=**35.3º** | [$11\bar{1}$] | [$1\bar{3}1$] | [$11\bar{1}$] | [$\bar{3}11$] |
| | | **58.5º** between domain walls [$11\bar{1}$] & [$1\bar{3}1$], or [$\bar{3}11$] 63.0º between domain walls [$1\bar{3}1$] & [$\bar{3}11$] | | | |
| | **(201)** surface, Euler angle θ=**63.4º** | [010] | | [$\bar{1}\bar{2}2$] | [$\bar{1}22$] |
| | | 48.2.6º between domain walls [010] & [$\bar{1}\bar{2}2$] or [$\bar{1}22$] **83.6º** between domain walls [$\bar{1}\bar{2}2$] & [$\bar{1}22$] | | | |

**Table S2** Survey of selected literature report on the structure of tetragonal CH$_3$NH$_3$PbI$_3$.

| Technique | Non-polar I4/mcm | Polar I4cm | Noncommittal |
|---|---|---|---|
| X-ray and neutron diffractions | *J. Mater. Chem. A*. **1**, 5628 (2013); *Chem. Commun.* **51**, 4180 (2015); *Adv. Funct. Mater.* **25**, 2378 (2015); *Sci. Rep.* **6**, 35685, (2016); *Sci. Rep.* **6**, 30680, (2016); | *Inorg. Chem.* **52**, 9019 (2013); *CrystEngComm.* **17**, 665 (2015). | *J. Mater. Chem. A*. **3**, 9298 (2015) |
| Optic SHG | *J. Phys. Chem. Lett.* **7**, 2412 (2016). | *Proc. Natl. Acad. Sci.* **114**, E5504 (2017). | |
| Macroscopic measurements | *Appl. Phys. Lett.* **106**, 173502 (2015); *ACS Energy Lett.* **1**, 142 (2016). | *Proc. Natl. Acad. Sci.* **114**, E5504 (2017). | |
| Microscopic PFM | *J. Phys. Chem. Lett.* **6**, 1408 (2015); *J. Phys. Chem. Lett.* **6**, 1155 (2015); | *J. Phys. Chem. Lett.* **5**, 3335 (2014); *J. Phys. Chem. Lett.* **6**, 1729 (2015); *J. Mater. Chem. A*. **3**, 7699 (2015); *J. Mater. Chem. A*. **4**, 756 (2016); *Nanoscale* **9**, 3806 (2017); *Energy Environ. Sci.* **10**, 950 (2017) | *J. Phys. Chem. C*. **120**, 5724 (2016); *Science Advances* **3**, e1602165 (2017) |
| TEM | | | *Nat. Commun.* **8**, 14547 (2017) |
| DFT and MD simulations | *J. Mater. Chem. A*. **3**, 8926 (2015) | *Nano Lett.* **14**, 2584 (2014); *Chem. Mater.* **26**, 6557 (2014); *J. Phys. Chem. Lett.* **6**, 693 (2015); *J. Phys. Chem. Lett.* **6**, 2223 (2015); *J. Phys. Chem. Lett.* **6**, 31 (2015); *J. Phys. Chem. Lett.* **6**, 1155 (2015) | |